# Two insulating phases in compressed Pr1-xCaxMnO3 thin films

M. Filippi, W. Prellier[1]

Laboratoire CRISMAT, CNRS UMR 6508, ENSICAEN,

6 Boulevard Maréchal Juin, 14050 Caen Cedex, France

P. Auban-Senzier, C.R. Pasquier

Laboratoire de Physique des Solides, UMR8502, CNRS - Université Paris-Sud,

F-91405 Orsay Cedex, France

The temperature dependent resistivity of two $Pr_{1-x}Ca_xMnO_3$ (x=0.5 and 0.4) thin films grown on $LaAlO_3$ has been studied as a function of hydrostatic pressure (up to 2.5 GPa) and magnetic field (up to 9T). Both samples show a monotonic decrease in the resistivity with an increase in pressure, corresponding to a change of -35% at 2.5 GPa. No pressure induced metal-to-insulator transition was observed in the temperature-dependent resistivity. The non-trivial interaction between high pressure and magnetic field reveals that the effect of pressure cannot be simply rescaled to that of a specific field, as has been reported for the corresponding bulk material. We propose an interpretation of the data based on phase separation, where two different insulating phases coexist: the charge ordered phase, which is sensitive to both magnetic field and pressure, and a second insulating phase that can be tuned by magnetic field. Such a result demonstrates that phase separation can be manipulated in thin films by independent application of magnetic field and/or external pressure.

---

[1] prellier@ensicaen.fr

Numerous studies have examined phase separation in manganites, which is now believed to be a key ingredient for their Colossal Magneto-Resistive (CMR) properties.[1,2] In the case of thin films, evidence for phase separation has been observed, even near room temperature in several systems such as $Pr_{0.5}Ca_{0.5}MnO_3$ and $Pr_{0.6}Ca_{0.4}MnO_3$,[3-5], $La_{0.5}Ca_{0.5}MnO_3$ [6] and $La_{2/3}Ca_{1/3}MnO_3$.[7,8] In the $Pr_{1-x}Ca_xMnO_3$ (PCMO) system, a direct study of the electronic nature of this phase separation by means of resistivity measurements is challenging, since the compound remains in a macroscopic insulating state even when exposed to a large magnetic field. For a thin film of $Pr_{0.5}Ca_{0.5}MnO_3$ composition, Prellier et al. have shown that a relatively low field can melt the charge ordered state, [4] despite the fact that the charge-ordered (CO) state is most stable at this composition.[9] Such behaviour results in different CMR properties for thin film samples when compared to bulk samples. Electronic inhomogeneities within the films have been invoked to explain these differences. More precisely, an unusual ferromagnetic phase with high Curie temperature is commonly observed in these films.[3-5] Thus, we expect a control and manipulation of this phase separation by external stimuli, such as magnetic field and pressure.

In this letter, we report a study of the phase separation in $Pr_{0.5}Ca_{0.5}MnO_3$ and $Pr_{0.6}Ca_{0.4}MnO_3$ thin films where compressive strain is provided by the $LaAlO_3$ (LAO) substrate. The resistivity was studied as a function of hydrostatic pressure and magnetic field. The results can be interpreted using a scenario involving phase separation where two different insulating phases are switched into a metallic one by external stimuli. The proposed scenario consists of two coexisting phases, a CO-antiferromagnetic insulator (CO-AFI) phase (that can be converted into a metallic one by pressure and magnetic field) and an Insulating-ferromagnetic (I-FM) phase (which is sensitive to magnetic field). The ability to tune the resistivity of PCMO by independent application of magnetic field and external pressure presents opportunities for devices application.

Thin films of PCMO with a typical thickness of 500Å were grown on (100)-oriented LAO substrate. The films were deposited at 720 °C by the pulsed laser deposition technique from a stoichiometric sintered pellet under a 200 mTorr atmosphere of flowing oxygen. After the deposition, the films were slowly cooled to room temperature under 300 mbar of oxygen. High hydrostatic pressure was provided by a clamped cell made of NiCrAl alloy with silicon oil as pressure transmitting medium. The pressure was measured at room temperature by a manganin gauge and the temperature by a Cu/constantan thermocouple, both located in the cell next to the measured film. The resistivity was measured with a direct current (1µA) using the standard four-contact method on films with typical dimensions (2x1mm$^2$).

Figure 1 shows the relative variation in the resistivity as a function of the hydrostatic pressure for both films at room temperature. The resistivity was measured both while increasing and releasing pressure from the clamped cell, and no irreversibility was found. The pressure reduces the resistivity of the films, with a slope of about 16% / GPa for the $Pr_{0.5}Ca_{0.5}MnO_3$ and 14% / GPa for the $Pr_{0.6}Ca_{0.4}MnO_3$. Under the maximum applied pressure of 2.5GPa, the resistance decreases by 40% and 35%, respectively in the two films. The effect of pressure on the resistivity and functional properties of CMR manganites has been studied widely in the past. In particular, resistivity versus pressure has been measured in $La_{0.88}Sr_{0.12}MnO_3$ thin films where it was found that resistivity decreases up to 0.81GPa, after which it increases rapidly. Here, a continuous and almost linear decrease of the resistivity is observed up to 2.5 GPa.

The combined effect of magnetic field (0,2,5,7 and 9T) and high pressure (0-2.5GPa) on the resistivity of $Pr_{0.5}Ca_{0.5}MnO_3$ is presented on Figure 2. Data are presented for the warming run and no thermal hysteresis was observed. Without magnetic field, the film is semiconducting, while a 7T applied magnetic field (not shown) induces a metal-to-insulator transition (MIT) around 150K, which leads to a large magnetoresistance value. This contrasts with bulk materials[11] and thicker films,[3] where only a moderate magnetoresistance without MIT has been observed. This difference is attributed to the thickness-dependence of the MIT, (as observed in PCMO films grown on SrTiO3[12]). An applied field of 9T further decreases and flattens off the resistivity. In addition, two distinct MIT (at around 80K and 140K) are observed, which may indicate that two different insulating phases are melted upon application of a magnetic field. In fact, microstructural studies previously reported [3] for thicker films (1200Å) have shown that the modulation vector of the CO state (q=0.35-0.4) is far from the expected value q=0.5, and a second insulating ferromagnetic phase was detected. This implies that the thin films are electronically more inhomogeneous than the bulk material, indicating the presence of a second insulating phase distinct from the CO one. It should be stressed that the microstructural studies performed by transmission electron microscopy did not reveal any chemical phase separation, indicating that the behaviour reported here is clearly coming from an electronic phase separation.

At 0T, under 2.5GPa, the sample is still semiconducting indicating that the pressure itself (at least up to 2.5GPa) is not enough to induce a macroscopic metallic state, even if a significant decrease in the resistivity is observed. However, the application of magnetic field under 2.5GPa flattens the resistivity (2T-5T-7T in the figure). At 9T, under 2.5GPa, the temperature dependence of the resistivity is flatter than in the 9T-0GPa curve, and the 140K MIT observed on the latter curve is suppressed. If we examine the data measured under pressure, the resistivity values around 140K are strongly reduced by the 9T field, which suggests the presence of a phase sensitive to field and pressure. In contrast, the second MIT (at 80K) induced by the application of 9T remains almost unaffected with and without pressure. It should be noted that the bulk phase diagram of PCMO in the pressure-temperature plane [13] indicates that the effect of pressure can be directly rescaled with the magnetic field, with 6T being equivalent to 1GPa. More recently,[14] it was reported that for Ca content of x=0.35 a pressure of 2GPa induces a MIT. For thin films, a recent Raman study revealed that pressure suppresses the CO component; however, the film always remains macroscopically an insulator.[15] Here, the effect of pressure cannot be rescaled simply to a specific field. It appears that the system is in a metastable state, where two insulating components are present, whose response to the temperature, magnetic field and pressure are different, which gives rise to the observed macroscopic behaviour.

Both insulating phases are sensitive to the magnetic field, suggesting their magnetic nature. According to a previous Raman experiment,[15] which suggested that the pressure suppresses the CO phase, we can identify for the 150K peak the melting of CO because this peak is flattened by pressure. The second peak (at 80K) can be identified as the stabilization of the I-FM phase previously observed3. However, the film used in the present study is thinner than those studied previously, and therefore its magnetic signal is completely overwhelmed by the substrate response. Consequently, the magnetic nature of these phases is not clear. Nonetheless, the strong magnetic field dependence of their transport properties suggests that both phases possess a magnetic order. The effect of pressure on the CO phase can be interpreted considering the double exchange theory, since the application of external pressure would increase the Mn-O-Mn angle toward 180°. Thus, the film can present a phase separation into 2 phases. The first one is I-FM and can be transitioned to metallic by a 7T field, but is not sensitive to the pressure. The second one is CO-AFI and can be tuned both by the magnetic and the hydrostatic pressure. Such a concept is further supported by a recent study of the magnetic properties of PCMO films probed by DC magnetization and magneto-optical Kerr Effect measurements, where a FM phase mixed with a AFI was observed.[16]

Similar experiments (figure 3) were performed on the $Pr_{0.4}Ca_{0.6}MnO_3$ film (a study of the effect of pressure on $Pr_{0.6}Ca_{0.4}MnO_3$ can also be found in Ref. 15, where the increased Ca content induces a slightly higher internal pressure. At ambient pressure (upper curves) the film shows a moderate magnetoresistance at 7T and no MIT is observed. This temperature behavior is typical of the charge ordering as seen from the anomaly in the resistivity. However, we do not observed a clear flex , as in bulk, most likely because the charge ordering is not fully developed here.[3,4] The lower curves are measured with an applied hydrostatic pressure of 2.5GPa, and either a or a 7T magnetic fields. The effect of pressure itself is much greater than that of the magnetic field. Nevertheless, if we compare the changes induced by 7T at 0GPa and 2.5GPa, it appears that a magnetic field of 7T has the same effect at ambient pressure and 2.5GPa. Therefore, the two effects seem not to be coupled; they are just additive. An estimate of the effect of pressure at a given temperature, where it is equivalent to magnetic field as in the bulk, leads to a very high value. In fact, if we consider the resistivity decrease induced by 2.5GPa at a given temperature (for example 200K), this value is 4 times larger than that of 7T, leading to a rough equivalence between 2.5GPa and 28T. However, an applied 7T magnetic field is enough to induce a MIT in the bulk. Hence, it is likely that the effects of pressure and magnetic field are not similar, and the magnetic field and the pressure appear to act on different non-interacting channels. Therefore, the idea of two insulating phases with different responses to field and pressure also can be invoked for this composition, even if in this case the external stimuli (7T, 2.5GPa) do not induce a macroscopic metallic state.

In conclusion, we have studied the pressure-dependent resistivity of $Pr_{0.5}Ca_{0.5}MnO_3$ and $Pr_{0.4}Ca_{0.6}MnO_3$ thin films deposited on $LaAlO_3$. The resistivity was found to be very sensitive to the external pressure, decreasing at a rate of 14-16% / GPa. We also explored the joint effect of pressure and magnetic field upon the resistivity. The results can be explained using a phase-separation scenario, where a CO insulating phase, which can be suppressed by magnetic field and pressure, coexists with a second insulating phase, which is sensitive to the field. We believe that these results will be useful in understanding the difference in properties between oxide thin films when compared to bulk materials.

We acknowledged financial support from the European project COMEPHS (NMP34-CT-2005-517039) and thank Dr. W.C. Sheets for careful reading of the manuscript.

Figure Captions

FIG. 1 Relative variation of the resistance at room temperature as a function of the pressure for the $Pr_{0.5}Ca_{0.5}MnO_3$ (circles) and $Pr_{0.6}Ca_{0.4}MnO_3$ (squares) thin films.

FIG. 2 Resistivity as a function of the temperature for the $Pr_{0.5}Ca_{0.5}MnO_3$ film taken under different magnetic fields and pressures (0T, 0GPa); (9T, 0GPa) (black curves) and (0T-2T-5T-7T-9T, 2.5GPa) (red curves) as indicated by the arrows.

FIG. 3 Resistivity as a function of the temperature for the $Pr_{0.6}Ca_{0.4}MnO_3$ film taken under different magnetic fields and pressures (0T, 0GPa); (7T, 0GPa); (0T, 2.5GPa); (7T, 2.5GPa) as indicated by the arrows.

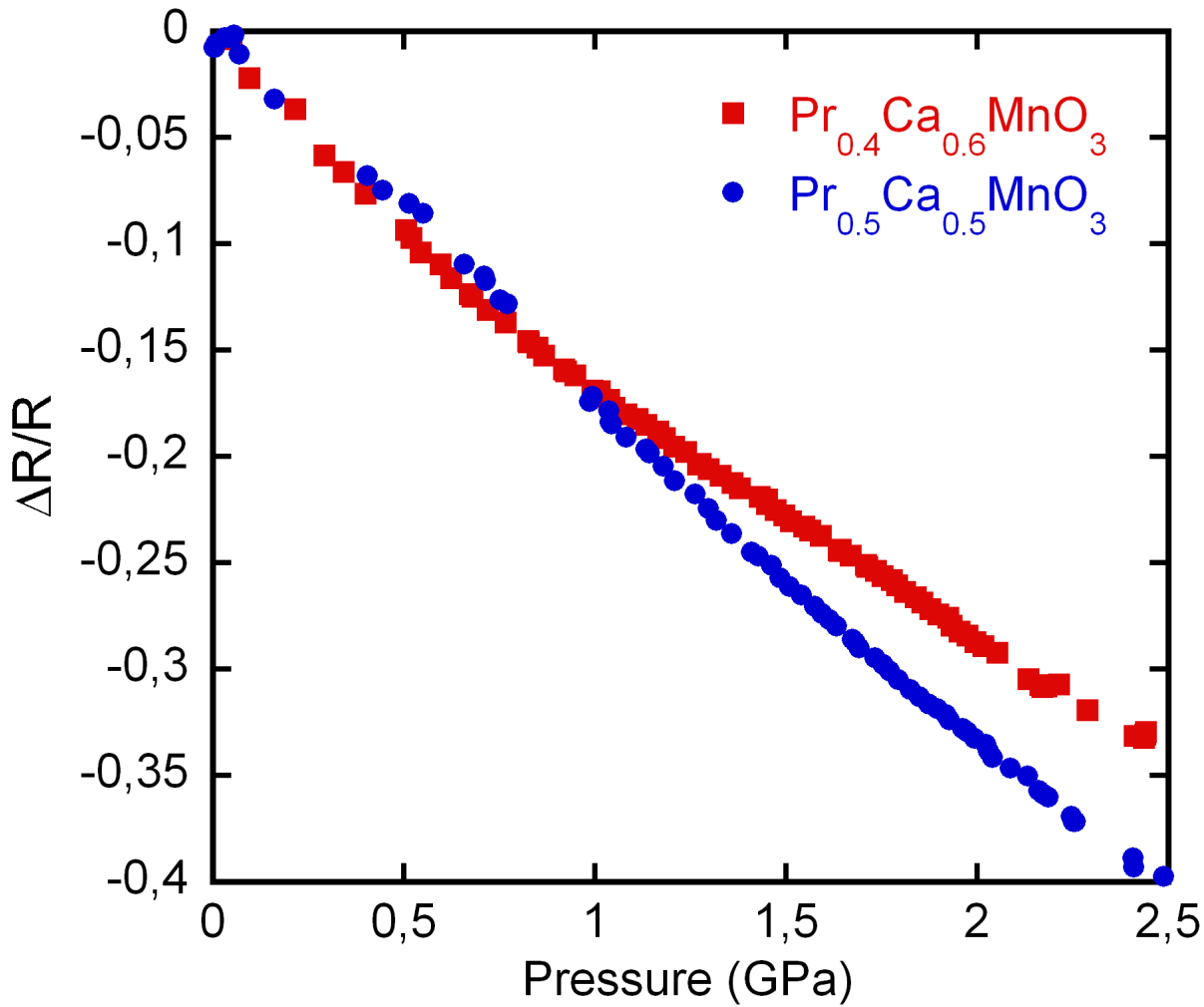

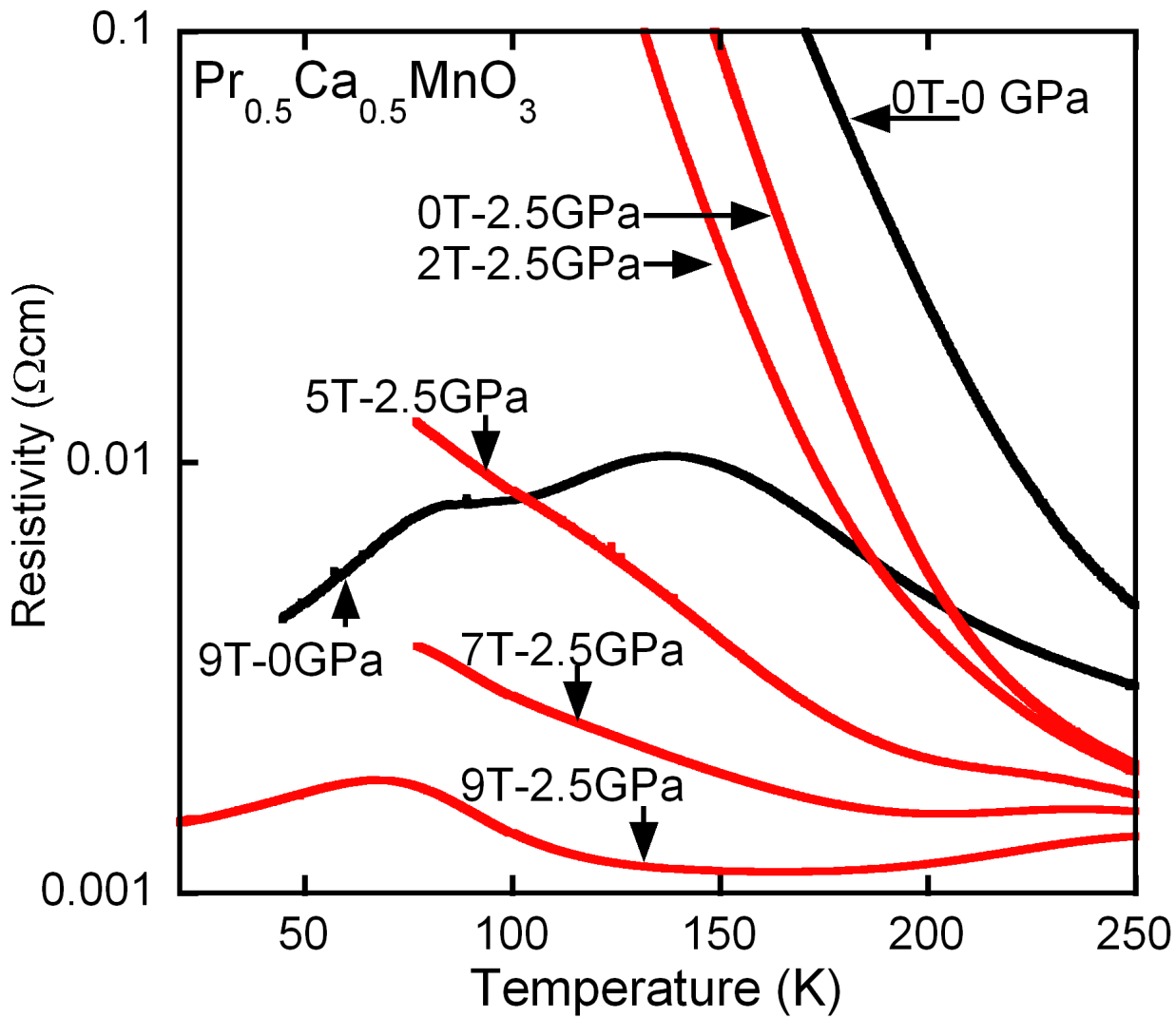

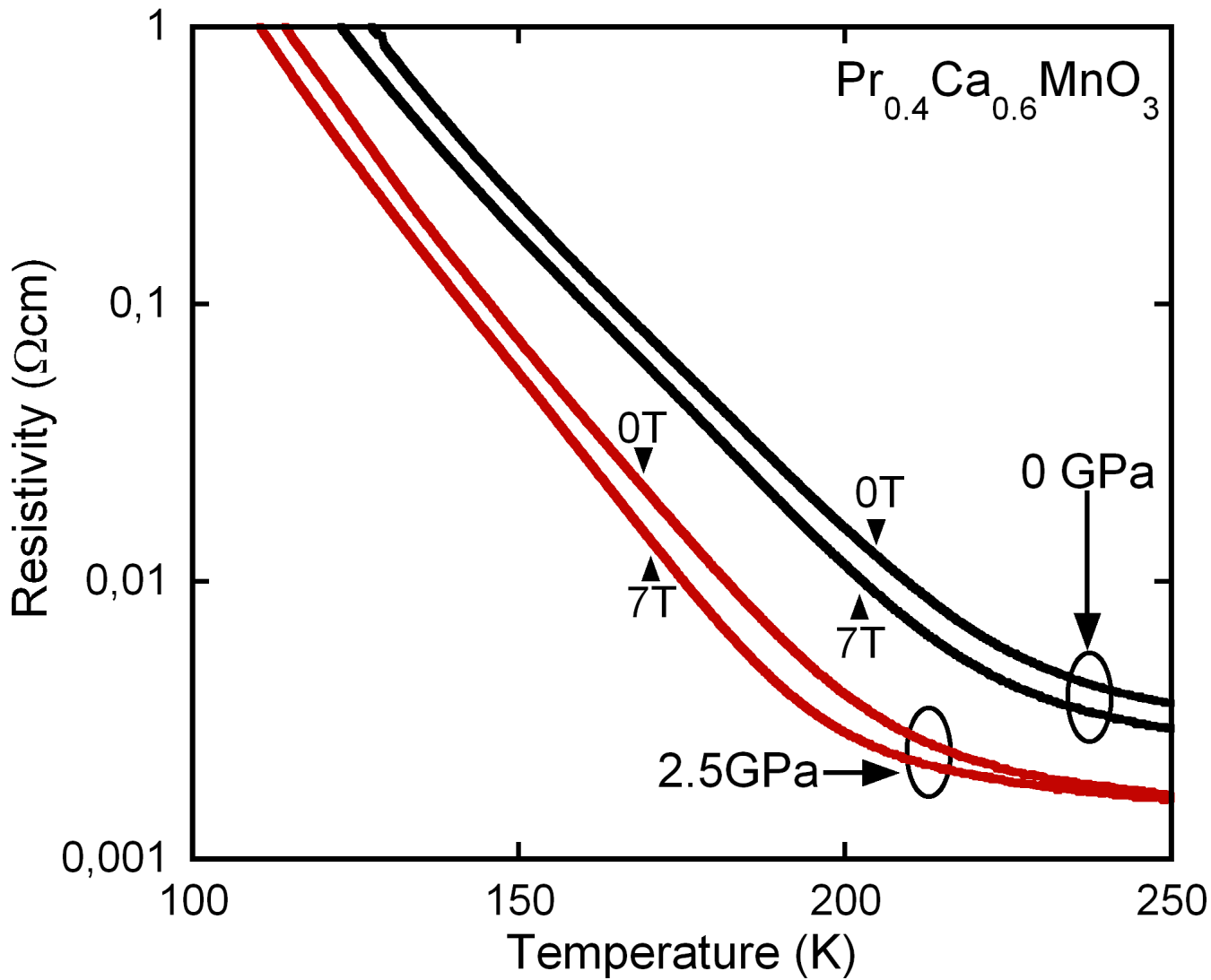